\documentclass[11pt]{article}
\usepackage[tikz]{bclogo}
\newcommand{\thetatwo}{\ensuremath{{\rm\Theta}_2^p}}

\usepackage[ascii]{inputenc}  %

\begin{document}
\sloppy
\title{Credimus}
\author{Edith Hemaspaandra\thanks{This work was done in part while
  on a sabbatical stay at ETH
  Z\"{u}rich's Department of Computer Science, generously supported by
  that department.}\\
Department of Computer Science\\
Rochester Institute of Technology\\
Rochester, NY 14623, USA \\
www.cs.rit.edu/$\sim$eh
\and Lane A.~Hemaspaandra$^*$\\
Department of Computer Science\\
University of Rochester\\
Rochester, NY 14627, USA \\
www.cs.rochester.edu/u/lane}
\date{November 1, 2017; revised June 1, 2019}

\maketitle

\begin{abstract}
  We believe that economic design 
and computational
  complexity---while already important to each other---%
should become even more important to each other with each passing year.
  But for that to happen, experts in on the one hand such areas as social
  choice, economics, and political science and on the other hand
  computational complexity will have to better understand each other's
  worldviews.

  This article, written by two
  complexity theorists who also work in computational social choice
  theory, 
focuses on 
one direction of that process by 
presenting a brief overview of how 
most computational complexity theorists 
view the
  world.  Although our immediate motivation is to make the lens through 
  which complexity theorists see the world be better understood by those 
  in the social sciences, we also feel that even within computer 
  science it is very important for nontheoreticians to understand how 
  theoreticians think, just as it is equally important 
within computer 
  science for theoreticians to understand how 
  nontheoreticians think.
\end{abstract}

\section{Introduction}
\label{s:intro}

Predictions are cheap.  Our cheap prediction is:
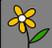
\begin{bclogo}[couleur=gray, arrondi=0.1]{}
Economic design and computational complexity should and will
in the future be even more deeply intertwined than they currently are.
\end{bclogo}

What is a bit less cheap is working to make predictions come true.  If
the prediction is broad or ambitious enough, doing that is often a
task beyond one paper, one lifetime, or one generation.

Nonetheless, in this article we will seek to make a small
contribution toward 
our predication's 
eventual
realization.
In particular, as complexity theorists who have for more than a decade
also been working in computational social choice theory, we have seen
first-hand how deeply important computational social choice theory and
computational complexity have been to each other.  And the ``to each
other'' there is not written casually.  As 
argued in a separate 
paper~\cite{hem:ctoappear:bffs},
the benefit of the interaction of those areas has been very much a
two-way street.

However, to increase the strength and quality of the interaction, and
to thus reap even more benefits and insights than are currently being
gained, a needed foundation will be \emph{mutual understanding}.
After all, even the different subareas of computer science have quite
different views of what computer science 
is about, and
sometimes it seems that computer scientists don't understand even each
other's worldviews.  

As complexity theorists, we can expertly address only one direction
in which 
explanation is needed:
trying to explain the---perhaps strange to those who are not complexity
theorists---way that complexity theorists tend to view the world.
We sincerely hope that the 
reciprocal directions will be addressed by 
appropriate experts from 
the many other disciplines whose practitioners are part of
the study of economic design.

Beyond that, we also embrace and somewhat generalize a 
hope that for the case of (computational) social choice
is expressed in the above article~\cite{hem:ctoappear:bffs}:
We hope that 
in time there will be a generation of researchers who are trained
through graduate programs that make students simultaneously expert in
computational complexity and one of the 
other 
disciplines underpinning economic design---researchers who in a single
person achieve a shared understanding of two areas.  But for now, most
researchers have as their core training one area, even if they do reach
out to work in---or work with experts in---another area.  And thus
we write this article to try to make as transparent as we can within
a few pages the crazy, yet (to our taste) just-right, way that complexity
theorists view the world.

The remainder of this article is organized as follows.
Section~\ref{s:need} further discusses the need for, and the
importance of improving, mutual understanding.  That section argues
that the way that complexity theorists view the world is rarely
understood even by the other areas of computer science, and that the
areas of computer science themselves are separated 
by huge cultural gaps.
Section~\ref{s:creed} presents what we feel is the heart of how
computational complexity theorists view the world, which is:

\begin{bclogo}[couleur=gray, arrondi=0.1]{}
  We as complexity theorists 
believe that there is a landscape
  of beautiful mathematical richness, coherence, and
  elegance---waiting for researchers to perceive it better and better
  with the passing of time---in which problems are grouped by their
  computational properties.
\end{bclogo}

\section{The Need for Creed: Why Understanding Each Other Is 
Hard yet Needed and Important}
\label{s:need}

In this section, we briefly mention some cultural chasms between
complexity and social choice---and even between complexity and other
areas of computer science---and suggest that shrinking or removing
those chasms is important: Understanding between collaborators 
is of great value to the collaboration.

\subsection{Spanning to Computational Social Choice, Economics, and Beyond}
In this section, we will focus on computational social choice,
as that is the particular facet of economic design that the authors 
are most familiar with.  

This paper
will not itself present the many ways that computational complexity
and computational social choice have interacted positively and in ways
that benefit \emph{both} areas.  As mentioned above, a separate
paper~\cite{hem:ctoappear:bffs} 
already makes that case, for example
pointing out: how computational social choice has populated with
problems such classes as the $\thetatwo$ level of the polynomial
hierarchy and $\rm NP^{PP}$; how computational social choice has provided the
first natural domain in which complexity theory's
search-versus-decision separation machinery could be applied; how that
application itself gives insight into how to best frame the
manipulative-attack definitions in computational social choice; how
complexity's ``join'' operation has been valuable in proving the
impossibility of obtaining certain impossibility results in
computational social choice theory; how the study of online control
yielded a completely new quantifier-alternation characterization of
coNP; and much more, such as how important
a role approximation, dichotomy results, and parameterized complexity
have played in computational social choice.  

It in fact is quite remarkable how strongly complexity has helped the
study of computational social choice, and is even more
remarkable---since this is the direction that might not have been
apparent beforehand---how strongly computational social choice has
helped the study of computational complexity theory.  And most
remarkable of all is that these results have 
usually been obtained 
by
researchers from one 
side,
although ones who were very interested in the other side.
This clearly is a pair of areas that already is showing very strong
content interactions and mutual benefits.  Think of how much more waits to be
seen and 
achieved when computational social choice theorists/social
choice theorists and complexity theorists deepen their understanding
of each other.

Some might worry about the ``computational'' in ``computational social
choice'' above---namely, worrying that computational social choice is
such a young area that no one is ``native'' to it.  We disagree.  
It is true that much of the key, early work on this area---as it was
emerging \emph{as} an area with a distinct identity%
{}---was done by researchers whose training was in operations
research, logic, artificial intelligence~(AI), 
theoretical computer
science, economics, social choice, political science, or mathematics.
But already a generation of students, now in their 20s and 30s, has
been trained whose thesis work was on computational social choice theory:
researchers whose 
``native'' area and 
identity is---despite the fact
that their thesis advisors view themselves as at their core part of
one of the older areas just listed---that they are computational
social choice theory researchers.  This is a very good development, and yet 
we are asking even more: Now that the area has its own identity, 
one can hope to grow researchers whose core training and identity
embraces both
that young area and the 
area of computational complexity.

\subsection{Spanning to 
Other Computer-Science Areas}
It is often discussed within computer science departments whether
computer science is even a coherent discipline.  After all, 
if one thinks about which other department the areas of computer science
feel kinship to, for theoreticians that generally would 
be mathematics, for systems people that generally would be electrical 
and computer engineering, for symbolic AI 
people that often would be one of 
brain and cognitive sciences, 
linguistics, philosophy, or 
psychology,
and for vision/robotics AI people that might be 
mechanical engineering, electrical and computer engineering, 
or visual science.  

The cultural differences are also stark.  For example, taking as our
examples the two subareas of computer science that are most strongly
represented in computational social choice research---AI and
theory---we have the following contrasts in culture.  Anonymous
submissions at the main conferences versus submissions with the
authors' identities open.  Intermediate feedback and rebuttals at the
main conferences versus no such round.  Authors' names generally being
ordered by contribution versus authors names always being listed
alphabetically.\footnote{One of us once asked a colleague, who 
at one point was the president of AAAI, whether 
he,
upon seeing 
a paper with a very large number of authors with them all
in alphabetical order, would really assume that Dr.~Aardvark 
had made the largest contribution.  The colleague looked back as if 
he'd been asked whether he really believed that $1+1=2$ and 
said that he of course 
would.  In fact, in the different area within computer science known as 
systems, there is a running semi-joke---that 
excellently corresponds with reality---that 
one can tell how theoretical a given systems conference is by 
looking at what portion of its papers list 
the authors in 
alphabetical order; in fact, there is a
very
funny 
joke-paper~\cite{app:j:alphabetical-order-in-systems} 
that quantifies this---more rigorously than 
the earlier part of this sentence does---to prove that the POPL 
conference is quite theoretical.}
Large hierarchical program committees versus 
small almost-flat program committees.
And that listing is not even 
mentioning the issue of the contrasting content of the areas, or 
their differing views on 
conference versus journal 
publication.

Almost any theoretical computer scientist will have stories of
how sharply his or her perspective has differed from those of 
his or her nontheory colleagues,
e.g., a nontheory colleague who firmly felt that 8x8 chess---not 
$N$x$N$ chess~\cite{sto:j:chess} but actual 8x8 chess---under 
the standard rules (which implicitly 
limit the length of any game) is a great 
example of \emph{asymptotic} complexity, and who advised 
the theoretician to go use 
Google to learn more about this.
We suspect that nontheory computer science faculty members could write 
quite similar sentences---with different examples---from their own 
points of view, regarding the things theory faculty members 
say.

So even the subareas of computer science have some gaps
between them
as to understanding, or at least have rather large 
agree-to-disagree differences.
Our hope is that, regarding the former, this short article may be 
helpful.  

We mention, however, that we do not agree that anything said above
shows
that computer science is not a coherent discipline.  To us,
and in this we are merely relating an important, much loved insight
that has been around in one form or another for many
decades~\cite{knu:c:algorithms-in-modern-mathematics-and-cs,har:b:algorithmics-1st-ed},
there is a unifying core to the field of computer
science: algorithmic thought (and the study of algorithms).
That core 
underpins AI, systems, and theory, and makes computer science an at
least 
decently coherent discipline.
\section{A Core Belief, and Its Expressions, Interpretations, and Implications}
\label{s:creed}

\subsection{A Core Belief}\label{ss:core}
We feel that a 
core view---in fact, \emph{the} core view---of complexity theorists
is the following
(phrased here both as a profession of belief and as a statement 
of what is 
believed).

\begin{bclogo}[couleur=gray, arrondi=0.1]{}
  [\textbf{Core Belief~~}%
We as complexity theorists believe that:]~There is a landscape
  of beautiful mathematical richness, coherence, and
  elegance---waiting for researchers to perceive it better and better
  with the passing of time---in which problems are grouped by their
  computational properties.

\end{bclogo}

If the subfield can be said to have a creed, this is it.  

By saying that complexity theorists feel this, we don't mean to
suggest that it is exclusive to them.  In a less computational vein,
the great mathematician Paul Erd\H{o}s spoke of ``The Book,'' which
holds the most elegant proof of each mathematical theorem.  He
famously said, ``You don't have to believe in God, but you should
believe in The Book,'' and surely viewed as moments of true joy those when a
proof so beautiful as to belong in the book was discovered.  And the
great computer scientist Edsger Dijkstra is traditionally 
credited\footnote{The quote is attributed to him in works of others as early
as 1993~\cite[p.~4]{hai:thesis:distributed-runtime-support}, though
attributing the quote to Dijkstra 
is disputed, as Michael Fellows published a very similar 
comment in a 1991 manuscript that appeared in a 1993 conference proceedings
and published the identical quotation in 1993 in 
a \emph{Computing Research News} article 
joint 
with Ian Parberry.}
with this lovely, insightful
comment: 
\begin{bclogo}[couleur=gray, arrondi=0.1]{}
  Computer Science is no more about computers than astronomy
is about telescopes.\ \mbox{---~E.~Dijk\-stra}
\end{bclogo}
\noindent
Though different people interpret that quotation in different ways,
we have always interpreted it to suggest almost precisely 
what our core belief is expressing.  
Indeed, 
the quotation's 
implicitly drawn parallel between
astronomy studying the structure of the universe and computer scientists
studying a similarly majestic structure is extremely powerful.
And things are made even more pointed in the 1991 version by Michael 
Fellows, which follows the same sentiment as that of the quotation 
with, ``There is an essential unity of mathematics and computer science.''

\subsection{The Heretics}
Having read Section~\ref{ss:core}, theoretical researchers from any
field may think, ``Well, duh!''  That is, they may think that the
core belief is obvious, and wonder who could possibly think anything else.

The answer is that quite a 
large portion of the field computer science
 thinks
something else.  This was most famously expressed in a 1999 ``Best
Practices 
Memo''~\cite{pat-sny-ull:j:best-practices-memo-experimentalism-as-ok-cra}
that was published in \emph{Computing Research News}, the newsletter
of a prestigious group, the Computing Research Association,
of over two hundred North American organizations
involved in computing research, including many universities.
To this day, that memo is on the Computing Research 
Association's web site as a best
practices memo 
(\cite{pat-sny-ull:url:best-practices-memo-experimentalism-as-ok-cra},
although 
there 
certainly has been strong pushback on some of its 
points~\cite{var:j:conferences-versus-journals,for:j:growup}).  
The most jump-off-the-page lines in that memo are
these:

\begin{quotation}
  ... experimentalists tend to conduct research that involves creating
  computational artifacts and assessing them. The ideas are embodied
  in the artifact, which could be a chip, circuit, computer, network,
  software, robot, etc. Artifacts can be compared to lab apparatus in
  other physical sciences or engineering in that they are a medium of
  experimentation. Unlike lab apparatus, however, computational
  artifacts embody the idea or concept as well as being a means to
  measure or observe it. Researchers test and measure the performance
  of the artifacts, evaluating their effectiveness at solving the
  target problem. A key research tradition is to share artifacts with
  other researchers to the greatest extent possible. Allowing one's
  colleagues to examine and use one's creation is a more intimate way
  of conveying one's ideas than journal publishing, and is seen to be
  more effective. For experimentalists conference publication is
  preferred to journal publication, and the premier conferences are
  generally more selective than the premier journals..\@. In these and
  other ways experimental research is at variance with conventional
  academic publication traditions.
\end{quotation}

Underlying this is a worldview that is very different than that of
most theoreticians.  The worldview is that software systems and
devices are often so complex that trying to theoretically
capture their behavior and properties is hopeless, 
and we instead need to experiment
on them to make observations.  For example, that view might suggest
that operating systems are so enormous and complex that we can't
really capture or understand precisely their behavior.  

Yet theoreticians think otherwise.  Theoreticians dream of a time when
essentially all programs---of any size---will have a rigorous,
formally specified relationship between their inputs and their
actions/outputs, and when we will seek to prove that the programs
satisfy those relationships (insofar as can be done without running
aground on undecidability issues).  Perhaps that time will be 
decades 
or centuries away for extremely complex programs,
but we believe it will come.  And in fact,
real progress---for example thanks to advances in automated
theorem-proving/automated reasoning---has been made in the past few
decades on verifying that even some quite large programming systems
meet their specifications.  

In brief, we don't think that because 
software systems are complex
one can only experiment on them as if they were 
great mysteries; rather, we think that, precisely because they are 
so complex, the field should increase its efforts to formally understand them,
including working on building the tools and 
techniques to underpin such an understanding.

To be fair to the above-quoted memo, it carefully had a very separate
coverage in which it described what theoreticians do.  But to many
theoreticians, viewing computing systems as too complex to
theoretically analyze---and more suitable for experimenting on---is
far too pessimistic, at least as a long-term view.

Is our Core Belief utterly optimistic?  Not purely so.  It is broadly
optimistic, in what it believes exists, though to be frank the
landscape it is speaking of is typically more about problems and
classes than about analyzing operating systems.  But embracing 
the Core Belief does not mean
that one must be delusional as to time frames.  For example, in Gasarch's
2012 P versus NP poll~\cite{gas:j:second-p-vs-np-poll}, only 53\%
percent of those polled felt that P versus NP would be resolved by the
year 2100.
3\% thought it would never be resolved, and 5\%
said they simply did not know when/if it will be resolved.

An astounding
92\%
of the polled theoreticians believe that
it will
eventually
be resolved, even though 
currently no path for imminently
resolving the question is in sight (see also the very 
grim possibility mentioned in the 1970s by Hartmanis and 
Hopcroft~\cite{har-hop:j:ind}: that the question might 
be independent of the axioms of set theory).\footnote{Note added:
  In Gasarch's just-published 2019 P versus NP
  poll~\cite{gas:j:third-p-vs-np-poll}, those figures of
  53\%/3\%/5\%/92\% have shifted to
  66\%/9\%/0\%/91\%.  Gasarch's comment on that is that
  ``[the shift from 53\% to 66\%]
  amazes me because, since 2012, there has been little
  (no?)~progress on resolving $\rm P =?\, NP$\@.
  Note that in 2019 more people [9\%] thought it would
  never be solved than in 2012 [3\%] or
  2002 [5\%]''~\cite{gas:j:third-p-vs-np-poll}.}
Theoreticians 
have generally taken to heart 
Sir Thomas Bacon's 1605 comment from \emph{The Advancement 
of Learning}:

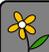
\begin{bclogo}[couleur=gray, arrondi=0.1]{}
  They are ill discoverers that think there is no land, when they can see nothing but sea.~---~Thomas Bacon
\end{bclogo}

\subsection{Landscape and Classification}
So what is this landscape that the Core Belief speaks of?
And how can we bring it into better focus?

\subsubsection{Axes and Granularity of Classification}
The landscape is one where each problem is located by its
classification in terms of various measures.  What is its (asymptotic,
of course) deterministic time cost?  What is its deterministic space
cost?  What are its nondeterministic costs?  Its costs in various
probabilistic models?  What about in nondeterministic models that
forbid ambiguity (i.e., that have at most 
one accepting path) or that polynomially
bound the ambiguity?  What about in quantum computing models 
and biocomputing models?
How well can the problem be---in various
senses---solved by heuristics or approximations?  What types of 
circuit families can capture the problem?  What types of 
interactive proof classes can capture the problem?  

And that is just a quick start to listing 
aspects of interest.  The number of interesting dimensions
along which problems can be classified is 
already large, and continues to grow
with time.  Our landscape is not a physical one, of course, but is
a rich world of mathematical classification.  

The granularity with which we group the ``locations'' in this world
itself is interesting.  Complexity theorists typically focus on
equivalence classes of problems, linked by some type of reduction.
For example, the NP-complete problems are all those problems that are
many-one, polynomial-time interreducible with the problem of testing
the satisfiability of boolean formulas.  One can think of the
NP-complete sets as an extremely important feature of the landscape.
Yet one can also view the landscape with an interest in other 
degrees of granularity.  The class of NP-Turing-complete
sets for example contains all the NP-complete sets, and may well
contain additional sets~\cite{lut-may:j:cook-karp}, since Turing
reductions are a more powerful reduction type than many-one
reductions.  Going in the other direction, the class of sets that are
polynomial-time isomorphic to boolean satisfiability may well be a
strict subset of the NP-complete sets, and it is known to be a strict
subset with probability one relative to a random
oracle~\cite{kur-mah-roy:j:prob1}.

Briefly put, complexity classes 
usually are defined by placing a bound on
some key resource, e.g.,
NP is the class of sets that can be accepted by polynomially
time-bounded nondeterministic computation.   Complexity 
classes in some sense are
upper bounds on some dimension of complexity.  
Reductions are
yardsticks by which sets can be compared.  If a set $A$ reduces to a
set $B$ by some standard reduction type, we view $A$ as being ``easier
or not too much harder'' than $B$, with the details depending on what
power the reduction itself possesses.  There are now a huge number of
intensely studied reduction types, capturing such notions as, just as
examples, the amount of time or space the reduction is itself allowed
to use; whether the reduction is a single query or multiple ones and
if the latter how they are used and whether they are sequential or
parallel; and to what extent the reduction itself can act
nondeterministically.  And completeness for complexity classes
combines a class with a reduction type, identifying those sets in
the class that are so powerful that every set in the class reduces to them
by the given reduction type.  In some sense, the completeness
equivalence class of a complexity class groups together those
problems, if any such problems 
exist (and some parts of the
landscape perhaps lack complete
sets~\cite{sip:c:complete-sets,gur:c:comp,har-hem:j:up,hem-jai-ver:j:up-turing}),
that distill the essence of the potential 
hardness of the class---they share the
same underlying computational challenge.  As such, they help
complexity theorists focus on what the source of a problem's 
complexity is.

The joyful obsession and life's work of complexity theorists 
is to better understand this landscape.  
This often is done though classifying where 
important problems---or groups of problems---fall.
Far more rarely yet vastly
more excitingly, complexity theorists 
find new relationships between
the different dimensions of classification,
e.g., by showing that every set in the polynomial hierarchy 
Turing reduces to probabilistic polynomial
time~\cite{tod:j:pp-ph}
or by showing the class of sets having interactive proofs 
is precisely deterministic polynomial 
space (PSPACE)~\cite{sha:j:ip}.

\subsubsection{Classification Is Done for Insights into the Landscape}
The Core Belief and the previous section should hint at a truth that
often is surprising to people who are not complexity theorists.  That 
truth
is that complexity theorists want to classify problems as part of the
ongoing attempts to better understand the landscape of problem
complexity.  And in particular, we are interested in doing that 
even for problems where the classifications we are trying to distinguish 
between don't in practice differ in what they say about how quickly a problem
can be solved.

For example, complexity theorists think that it is a rather big deal
whether a problem---if it is an interesting one, such as about
logic---is complete for double exponential time versus for example
being complete for triple exponential space.  This isn't because we
think that complete problems for double exponential time are 
going to be easy to quickly solve.  It is because we want to clarify
where interesting problems fall in the landscape. 

Looking at the other extreme, there is a huge amount of research into
complexity classes (such as certain uniform circuit classes and
logarithmic-space classes) all of which are contained in deterministic
polynomial time.  Yet to most people, deterministic polynomial time
already is the promised land as to computational cost.  Nonetheless,
smaller classes are intensely studied, to better understand the rich
world of complexities that exist there, and which problems have which
complexities, although in fairness we should mention that some of this
type of study is also motivated by the issue of whether the problem
can or cannot be parallelized~\cite{gre-hov-ruz:b:limits}.

But the real kicker here is that even if SAT solvers 
turn out to be
able to do stunningly well on NP-complete problems, complexity
theorists still will view the notion of NP-completeness as being of
fundamental importance to the landscape.
This is not because we don't
care about how well heuristics can do---that too is a dimension of the
landscape, and thus something on which rigorous results are important and
welcome---but rather 
we think that the notion of NP-completeness
itself is one of the greatest beauties of the 
landscape, and is natural and compelling in so very many ways.\footnote{%
This article is not on the subject of how well heuristics can do on
NP-complete problems, or the strengths and limitations of SAT solvers.
On one hand, there are theoretical results showing that
polynomial-time heuristics cannot have a subexponentially dense set of
errors on any NP-hard problem unless the polynomial hierarchy
collapses.  And if someone says they have a SAT solver that works on
any collection of NP problems they ever have encountered, it is
interesting to point out to them that factoring numbers that are the
product of two large primes can be turned into a SAT problem, and so
their amazing SAT solver should be able to break RSA and make them
rich...~yet no one has yet been able to 
make that work.  On
the other hand, SAT solvers undeniably 
do perform remarkably well on a great range of
data sets.  For discussion of 
most of the issues just mentioned, and how they can be at
least partially reconciled, see for example the article by
Hemaspaandra and
Williams~\cite{hem-wil:j:heuristic-algorithms-correctness-frequency}.}

To take as an example one of the most beautiful examples of how
profound the issue is of whether NP-complete sets belong to P, i.e.,
whether $\rm P = NP$, we mention that a not widely known paper by
Hartmanis and Yesha~\cite{har-yes:j:computation} is in effect 
showing
that whether humans can be perfectly
replaced by machines in the task of finding and
presenting particular-sized mathematical proofs
of theorems---loosely put, the issue of whether 
humans have any chance of having any special creativity 
and importance in achieving mathematical proofs---can
be characterized 
by the outcome of such basic landscape 
questions as whether P and NP differ, and whether 
P and PSPACE differ.

\section{Conclusion}
To end as we started, we believe that economic design and
computational complexity should become even more important to each
other with each passing year, but that an improved mutual
understanding of the areas' worldviews is important in making that
happen.  In that spirit, this article sets out the optimistic
worldview that we believe is held by most computational complexity
theorists.  And the most central part of that worldview is that
\emph{we as complexity theorists believe that there is a landscape of
  beautiful mathematical richness, coherence, and elegance---waiting
  for researchers to perceive it better and better with the passing of
  time---in which problems are grouped by their computational
  properties.}

That is not to say that we believe that the greatest open issues
within that landscape will be resolved within our lifetimes.  But we
believe that---just as that landscape has already been seen to have
utter surprises in what it says regarding language
theory~\cite{sze:j:csls,imm:j:csls}, interactive
proofs~\cite{for-kar-lun-nis:j:ip,sha:j:ip},
branching programs 
and safe-storage machines~\cite{bar:j:branching,cai-fur:j:bottleneck}, 
approximation~\cite{aro-lun-mot-sud-sze:j:proof}, the power 
and lack of power
of probabilistic
computation~\cite{nis-wig:j:hard,imp-wig:c:p-is-bpp-if-e-needs-esize-circuits,tod:j:pp-ph},
and much more---the landscape contains countless more surprises and advances
that will be reached in years, in decades, and in centuries, 
and we believe that many of them will be in the important, 
rapidly growing areas 
at the intersection of economic design and computational complexity.

\section*{Acknowledgments}
We thank Juraj Hromkovi\v{c} for hosting the sabbatical visit during which this
work was in part done.
We thank William S. Zwicker for helpful comments and suggestions.
\bibliographystyle{alpha}
%
\newcommand{\etalchar}[1]{$^{#1}$}

\end{document}